\documentclass[useAMS,usenatbib]{mn2e}
\usepackage{rotating}
\usepackage{journals}
\usepackage{graphics,graphicx}

\newcommand\msun {M$_{\odot}$}
\newcommand\gtrsim{\mathrel{\hbox{\rlap{\hbox{\lower4pt\hbox{$\sim$}}}\hbox{$>$}}}}
\def\approxgt{\ifmmode \rlap{$>$}{}_{{}_{{}_{\textstyle\sim}}} \else%
$\rlap{$>$}{}_{{}_{{}_{\textstyle\sim}}}$\fi} 
\def\approxlt{\ifmmode \rlap{$<$}{}_{{}_{{}_{\textstyle\sim}}} \else%
$\rlap{$<$}{}_{{}_{{}_{\textstyle\sim}}}$\fi}

\def\arcmin{\hbox{$^\prime$}}
\def\arcsec{\hbox{$^{\prime\prime}$}}

\def\flx{erg cm$^{-2}$ s$^{-1}$}
\def\lum{erg s$^{-1}$}

\def\chan{{\it Chandra}}
\def\src{CXO~J122518.6+144545}

\LARGE \normalsize \title[A bright off-nuclear X-ray source]{A bright
  off-nuclear X-ray source: a type IIn supernova, a bright ULX or a recoiling
  super-massive black hole in \src}
\author[Jonker et al.]  {P.G.~Jonker$^{1,2,3}$,
M.A.P.~Torres$^2$,  A.C.~Fabian$^4$, M.~Heida$^5$, G.~Miniutti$^6$, D.~Pooley$^7$\\ 
$^1$SRON, Netherlands Institute for Space Research,
Sorbonnelaan 2, 3584~CA, Utrecht, The Netherlands\\ 
$^2$Harvard--Smithsonian Center for Astrophysics, 60 Garden Street, Cambridge, MA~02138, U.S.A.\\
$^3$Department of Astrophysics, IMAPP, Radboud University Nijmegen, PO Box 9010, NL-6500 GL Nijmegen, the Netherlands\\
$^4$Institute of Astronomy, Madingley Road, Cambridge CB3 0HA\\
$^5$Sterrekundig Instituut Utrecht, P.O. Box 80 000, 3584 TA Utrecht, The Netherlands \\
$^6$LAEX, Centro de Astrobiologia (CSIC-INTA) LAEFF, PO Box 78, E-28691 Villanueva de la C\~anada, Madrid, Spain\\
$^7$Astronomy Department, University of Wisconsin, 475 North Charter Street, Madison, WI 53706, U.S.A.\\
}
\begin{document}

\maketitle

\begin{abstract} \noindent In this Paper we report the discovery of
  \src; a peculiar X-ray source with a position
  3.6$\pm$0.2\arcsec\,off-nuclear from an SDSS DR7 z=0.0447 galaxy.
  The 3.6\arcsec\, offset corresponds to 3.2 kpc at the distance of
  the galaxy.  The 0.3--8 keV X-ray flux of this source is 5$\times
  10^{-14}$ \flx\, and its 0.3--8 keV luminosity is 2.2$\times
  10^{41}$ \lum\, (2.7$\times 10^{41}$ \lum; 0.5--10 keV) assuming the
  source belongs to the associated galaxy. We find a candidate optical
  counterpart in archival HST/ACS g$^\prime$--band observations of the
  field containing the galaxy obtained on June 16, 2003. The observed
  magnitude of g$^\prime=26.4 \pm 0.1$ corresponds to an absolute magnitude of
  $-10.1$. We discuss the possible nature of the X-ray source
  and its associated candidate optical counterpart and conclude that
  the source is either a very blue type IIn supernova, a ULX with
  a very bright optical counterpart or a recoiling super-massive
  black hole. \end{abstract}


\section{Introduction} 

Off-nuclear ultra--luminous X--ray sources (ULXs) have been found in different galaxies. It has been
found that many of the ULXs are associated with active star--forming
regions (e.g.~in the Antennae, \citealt{2003ApJ...591..843F}), optical
emission line nebulae (e.g.~\citealt{2006IAUS..230..293P}) and radio
halos (e.g.~\citealt{2005ApJ...623L.109M};
\citealt{2007ApJ...666...79L}). 

In addition, six off-nuclear sources
with ${\rm L_x \gtrsim 10^{41}}$\lum\, have been found (i.e.~in the Cartwheel galaxy, \citealt{2006MNRAS.373.1627W}; M~82~X-1,
\citealt{2009ApJ...696.1712F} and \citealt{2003ApJ...586L..61S}; in
NGC~2276, \citealt{2004ApJ...604..653D}; ESO~243-49,
\citealt{2009Natur.460...73F}; in NGC~5775,
\citealt{2008MNRAS.390...59L} and M101~ULX-1, \citealt{2004ApJ...617L..49K}).  They are called hyper-luminous X-ray sources.
The optical counterpart has been found and studied in only a sub-sample of these (M101~ULX-1, \citealt{2004AJ....128.2783K}, \citealt{2009ApJ...704.1628L}; ESO~243-49, \citealt{2009arXiv0910.1356S}; and possibly M~82~X-1, \citealt{2004MNRAS.348L..28K}).

An important question in the work on the ULXs is whether the emission is isotropic
or not. If beaming is significant (\citealt{2001ApJ...552L.109K}) an
intermediate--mass black hole (IMBH) or super-massive black hole (SMBH) is not needed to explain sources with luminosities up to $\approx 10^{40}$\lum. For several sources, ionizing
luminosities derived from the optical emission lines and the resolved
radio bubbles indicate that beaming is not important
(\citealt{2003RMxAC..15..197P}; \citealt{2005ApJ...623L.109M};
\citealt{2007ApJ...666...79L}) and hence these sources are good
candidates for IMBHs. For the sample of hyper-luminous ULXs, their
X-ray luminosity seems too high for a stellar mass black hole even in the
presence of some beaming. For their classification there are currently three possible scenarios: very bright type IIn supernovae, IMBHs and recoiling SMBHs. We will briefly introduce these options.

The X-ray brightest supernovae are of type IIn (\citealt{2003LNP...598...91I}). Therefore, in principle, they could be responsible for a sub-sample of the very bright ULXs. Time variability in the X-ray as well as in the optical band can be used to constrain this possibility.

In the cold dark matter (${\mathrm \Lambda CDM}$) cosmological scenario
current (z=0) galaxies are the product of hierarchical mergers of
smaller galaxies. These smaller building blocks also host black holes
in their centers (\citealt{1995ARA&A..33..581K}). Evidence for this
comes from the observed M--$\sigma$ relation (\citealt{1998AJ....115.2285M}; \citealt{2000ApJ...539L...9F}). Furthermore, several so
called dual Active Galactic Nuclei (AGNs) have been found
(cf.~\citealt{2009ApJ...702L..82C} and references therein). After a
galaxy merger, the two black holes will eventually merge
as well (see \citealt{1980Natur.287..307B}). 

Recent fully relativistic numerical simulations allow for the
calculation of the linear angular momentum that is transported by
gravitational wave radiation during the final plunge in the black hole
-- black hole merger (\citealt{2005PhRvL..95l1101P}). The transport of
linear angular momentum acts as a kick on the newly formed black hole
merger product: gravitational wave recoil.  This effect had been
estimated before by for instance \citet{1973ApJ...183..657B} and
\citet{1989ComAp..14..165R}. When a black hole merger takes place in the presence of an accretion
disc, the recoiling black hole will take along the part of the nuclear
star cluster and accretion disc that falls within its gravitational
influence area (${\rm R_{infl}\approx 0.3 M_8 / v_{1000}^2\,pc}$; where $M_8$ is the mass in
units of $10^8$\msun\,and $v_{1000}$ is the velocity in units of 1000
km s$^{-1}$; Bonning, Shields \& Salviander 2007). Binary SMBHs
surrounded by an accretion disc will have emptied the inner portion
leaving a gap in the disc. Upon a kick this gap will refill on a short
timescale (\citealt{2007PhRvL..99d1103L}) and AGN activity will
resume.  Subsequently, such recoiling black holes may become visible
as off--nuclear AGNs (\citealt{2007ApJ...666L..13B};
\citealt{2008ApJ...687L..57V}; \citealt{2009ApJ...691.1050F}). The disc mass will allow for the AGN activity to last
tens of millions of years (cf.~\citealt{2007ApJ...666L..13B}).

The source SDSS~J092712.65+294344.0 has been proposed as a recoiling
massive black hole (\citealt{2008ApJ...678L..81K}). However, different
interpretations as a binary black hole (\citealt{2009ApJ...697..288B};
\citealt{2009MNRAS.398L..73D}) or a chance alignment with a more
distant source in the same cluster have also been proposed
(\citealt{2009ApJ...696.1367S}). Similarly, SDSS~J105041.35+345631.3
and SDSS~J153636.22+044127.0 have been suggested as recoiling SMBH
candidates (\citealt{2009ApJ...707..936S};
\citealt{2009Natur.458...53B}, respectively). The main difference between the IMBH and the SMBH scenario for the hyper-luminous sources lies in their optical properties. The recoiling SMBHs should carry with it the broad line region. Therefore, they should contain broad lines in their optical spectra whereas this is not the case for the IMBH scenario.

In this Manuscript we discuss the nature of another hyper-luminous off-nuclear
X-ray source: \src.

\section{Observations, analysis and results} 

\subsection{{\it Chandra} X-ray observation} 

In order to search for bright off-nuclear X-ray sources, we have
selected galaxies from the Sloan Digital Sky Survey (SDSS) data
release (DR) number 7.  We have cross-correlated this database with
the \chan\, source catalog in order to search for X-ray active
sources. Out of the resulting matches we have selected those for which
the \chan\, detection does not coincide with the position of the
galaxy center. We selected sources where the distance between the
center of the galaxy as determined by the SDSS-DR7 and the X-ray
position is less than 10\arcsec\, but larger than
2\arcsec. Furthermore, the error on the X-ray position had to be less
than 2\arcsec.  Next, we determined the redshift of the galaxy
(photometric or spectroscopic) from the SDSS database and we kept
sources with a luminosity larger than 1$\times 10^{40}$ erg s$^{-1}$
using the flux measured in the {\it Chandra} source catalog. Finally,
we have plotted the X-ray position on the SDSS $r^\prime$-band image
to visually verify the resulting sources. Below, we report on the
X-ray source CXO~J122518.6+144545.

\begin{figure}
\includegraphics[width=8cm,angle=0]{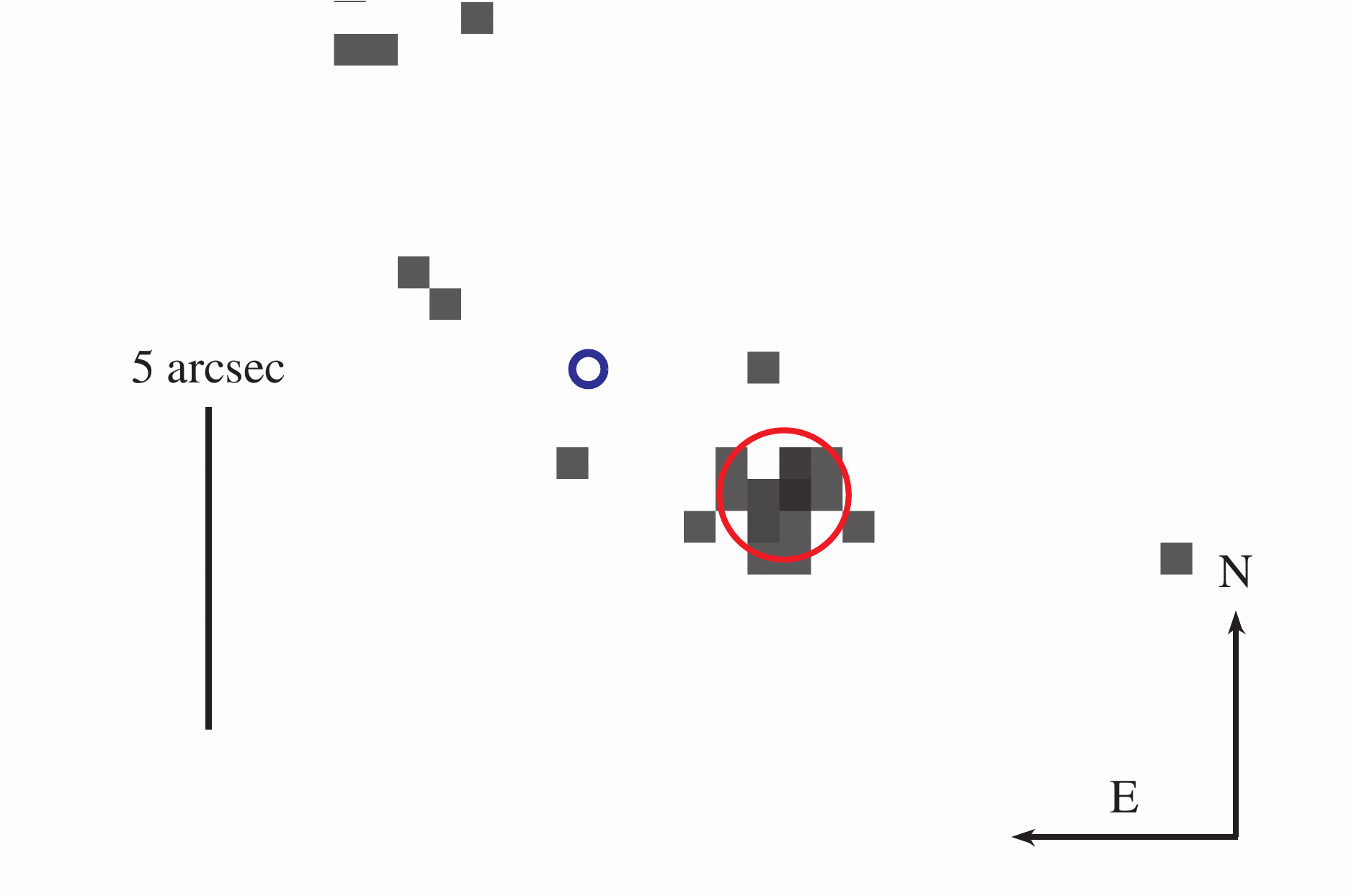}
\caption{Zoom in around the \chan\, ACIS-S3 position of the off-nuclear source 
(indicated with the red (larger) circle with a radius of 1\arcsec). The blue (smaller, thick line) circle indicates the SDSS DR7 position of the centre of the galaxy, its radius is
0.25\arcsec.}\label{ima} \end{figure}

\begin{figure}
\includegraphics[width=8cm,angle=0]{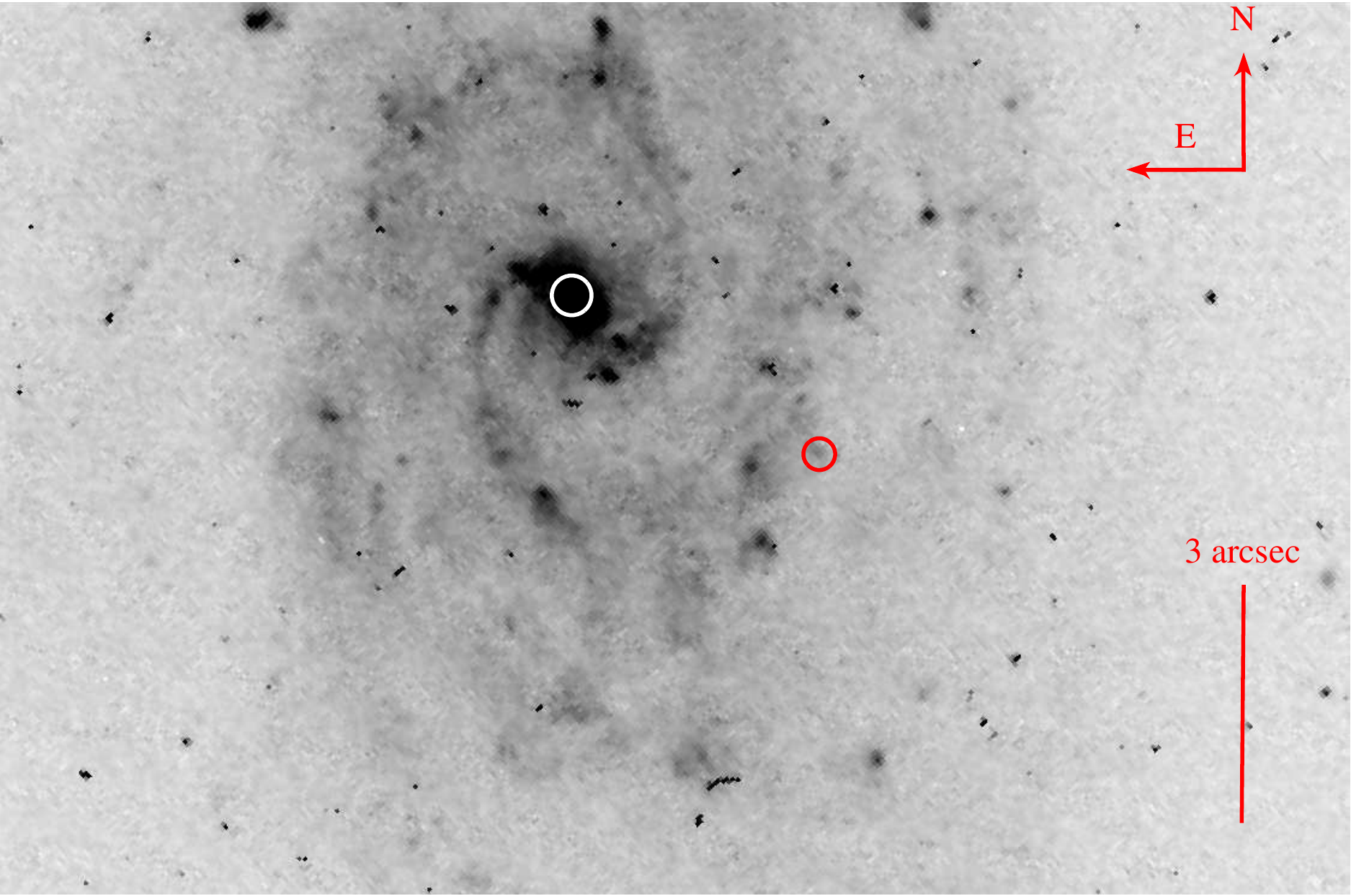}
\caption{HST/ACS $g^\prime$--band image with an exposure  time of 750~s revealing
the (cluster of) stars in the \chan\, error circle.   The grey circle (red in the colour version, South-West of the image centre) is at the
\chan\, position and the radius of 0.2\arcsec\,is equal to the overall  astrometric
error in the position of the X-ray source. The white circle indicates the SDSS DR7
position of the centre of the galaxy, the radius is 0.25\arcsec. The figure is made from the combined, 750~s-long drizzled HST image. }\label{hstima}
\end{figure}

This source has a position 3.6\arcsec\, off-nuclear from a galaxy
identified in the SDSS DR7 as having a redshift $z$=0.0447 (see
Fig.~\ref{ima} and \ref{hstima}).  The SDSS DR7 measured centre of
this galaxy is at Right Ascencion (R.A.; J2000)=12:25:18.860 (186.328583 in
decimal degrees) and the Declination (Dec; J2000)=14:45:47.704 (14.763251 in
decimal degrees). We have retrieved the X-ray observation with
observation ID 8055 from the \chan\, archive and reprocessed the
events with calibrations available in CALDB version 4.1.3 using the
version 4.1.2 of the \chan\, X-ray center {\sc ciao} tools. The
exposure time for the observation with ID 8055 is 5093~s. The source
is detected 1.7\arcmin\,off-axis on the ACIS S3 CCD.

Using {\sc wavdetect} we detect another source 84.3\arcsec\,away from
\src. This source is the brightest detected X-ray source and it can be identified with a bright point source in the SDSS $g^\prime$- and $r^\prime$-band images ($g^\prime$=20.10$\pm$0.02 and $r^\prime$=19.25$\pm$0.01). We use its accurate SDSS DR7 $r^\prime$-band position of R.A.=12:25:19.451 (186.33104668) and
Dec=14:47:09.262 (14.78590606) to determine a boresight correction for the \chan\,
observation. The optical positional accuracy depends on the
localization uncertainty which for stars with $r^\prime<20$ is negligible (\citealt{2003AJ....125.1559P}) and  on the statistical and systematic uncertainty in tying
the SDSS $r^\prime$-band field to the ICRS reference frame. We use the
conservative values provided by \citet{2003AJ....125.1559P} (see
Table~\ref{log}). 

\begin{table}
\caption{Sources of uncertainty used for the astrometry.}
\label{log}
\begin{center}
\begin{tabular}{cc}
\hline
Source & Uncertainty \\
\hline
Uncertainty tying SDSS frame to ICRS & 30 milli-arcsec$^{**}$\\
Uncertainty tying SDSS frame to ICRS & 60 milli-arcsec$^*$\\
X-ray localization uncertainty CXO source & 0.15 arcsec$^*$ \\
Remaining X-ray bore sight uncertainty & 0.12 arcsec$^*$ \\

\hline
\end{tabular}		      
\end{center}
{\footnotesize $^*$ statistical uncertainty}. \\
{\footnotesize $^{**}$ systematic uncertainty}. \\
\end{table}

Using this accurate $r^\prime$-band position, we determine a
bore-sight correction of dRA:0.03$\pm$0.08\arcsec\,and
dDEC:-0.23$\pm$0.08\arcsec. The error of 0.08\arcsec\, on each coordinate stems from the uncertainty in localizing the X-ray bore-sight correction source. After applying this boresight correction using the tool {\sc wcs\_update} we obtain a best fit position for the
off-nuclear X-ray source \src\, of R.A.=12:25:18.650(9) [186.32771(3)
degrees] and Dec=14:45:45.76(8) [14.76271(2) degrees] using {\sc
  wavdetect} where the number in-between brackets denotes the 68 per
cent confidence uncertainty in the last digit. These errors are due to
the error in the source localization on the CCD alone.

The final uncertainty on the position of the off-nuclear X-ray source is the
quadratic sum of the centroiding uncertainty of the X-ray source, the
uncertainty in the X-ray bore-sight correction and the uncertainty in
the astrometry of the $r^\prime$-band image (see Table~\ref{log}). The
overall astrometric uncertainty on the X-ray source position is
0.19\arcsec\, which we round-off to 0.2\arcsec.

Due to larger centroiding uncertainties in the positions of fainter
X-ray sources the association of such fainter sources with a candidate
optical counterpart is less certain. Therefore, we prefer to use the
bore-sight correction determined using only the one bright X-ray source
mentioned above. This does imply that we ignore any uncertainty in the roll
angle of the satellite, which could introduce a small effect in the
bore-sight correction.  Since we also improve the astrometry HST
images using the SDSS $r^\prime$-band (see below) we could use
astrometry relative to the SDSS $r^\prime$, however, the contribution
of linking the SDSS $r^\prime$ frame to the ICRS astrometric standard
frame to the error budget is small, hence we prefer to use the
absolute astrometric solution.

We have selected a circular region of 6 pixel ($\approx$3\arcsec)
radius centered on the source position to extract the source counts in
the energy range of 0.3--8 keV. We limited the radius to exclude the
centre of the galaxy.  Similarly, we have used a circular region with
a radius of 80 pixels away from any source but on the same S3 CCD
to extract background counts. We have made redistribution and
auxilliary response matrices for the source and background region
separately.

The net number of background subtracted source counts is 22. The
predicted number of background source photons is 4--5. Standard
Poisson statistics shows that this is a very significant detection
with a chance of less than 1$\times 10^{-8}$ to be due to a
fluctuation in the background. Using {\sc xspec} version 11.3.2p
(\citealt{ar1996}) we have fitted the spectrum of \src\ using Cash
statistics (\citealt{1979ApJ...228..939C}) modified to account for the
subtraction of background counts, the so called
W-statistics\footnote{see
  http://heasarc.gsfc.nasa.gov/docs/xanadu/xspec/manual/}. We have
used an absorbed power-law model ({\sc pegpwrlw}) to
describe the data.

Due to the relatively low number of counts we fix the interstellar
extinction during the fit to 2.8$\times 10^{20}$ cm$^{-2}$; the
Galactic foreground ${\rm N_H}$ in the direction of the source found by \citet{1990ARA&A..28..215D}.
The power-law index and normalisation were allowed to float. The
errors on the parameters are substantial due to the low number of
counts; we obtain a power-law index of 0.9$\pm$0.3 and an unabsorbed
flux of $(5.4\pm1.6)\times 10^{-14}$ \flx\, in the range 0.3--8 keV.
The errors are at the 68 per cent confidence level. If we fix the
power-law index to 1.9, such as found often for AGN, the extinction is
$(5\pm2)\times 10^{21}$ cm$^{-2}$ implying a significant amount of
extinction above the Galactic extinction in the direction of the
source.

Using standard cosmology the redshift converts to a distance of 182.6
Mpc, which makes the X-ray luminosity in the range 0.3--8 keV
${\mathrm L_X=2.2\times 10^{41}}$\lum, for comparison the 0.5--10 keV
luminosity is ${\mathrm L_X=2.7\times 10^{41}}$\lum.  We searched for
variability in the rate of arrival of the photons but we found none.
Despite the appearance of two flare-like features in the lightcurve, a
Kolmogorov--Smirnov (\citealt{prteve1992}) test showed that the
probability that the data are consistent with the null-hypothesis of a
constant photon arrival rate is 61 per cent.

\subsection{{\it HST} ACS observations }

We have analysed archival Hubble Space Telescope (HST) data obtained
with the Advanced Camera for Surveys (ACS) on June 16, 2003 (MJD
52806), program GO-9401 and with the Wide Field Planetary Camera 2 (WFPC2) on February 15, 2008 (MJD 54511), program GO~11083. The ACS/WFC HST observations consist of $2
\times 560$~s and $1 \times 90$~s exposures in the F850LP (SDSS $z^\prime$-band) filter
and $2 \times 375$~s exposures in the F475W (SDSS $g^\prime$-band)
filter. The WFPC2 observations consist of 3 $\times$ 700~s exposures in the F300W filter ($\lambda=2919.8\AA; \Delta \lambda=740.2$\AA). 

The photometry of the ACS images was performed using the ACS module in the software package {\sc dolphot} (version 1.1)\footnote{http://purcell.as.arizona.edu/dolphot/}.  Following the
{\sc dolphot}/ACS User's Guide, the images were processed by masking
all bad pixels using the {\sc acsmask} task and the multi-extension
FITS files were split into single chip images using the {\sc
  splitgroups} task before performing photometry. Finally the sky
background for each chip was calculated with the {\sc calcsky}
task. We run {\sc dolphot} on both bias and flat-field corrected
{\tt{flt}} images and on cosmic ray cleaned and flat-field corrected
({\tt{clr}}) archival images. We tested different sets of parameters
for the photometric measurements. These tests include the setting
recommended in the {\sc dolphot}/ACS User's Guide and the settings
described in \citet{2009ApJS..183...67D}.  The different parameter
sets provide results in agreement within the photometric errors. We
report here the measurements derived using the parameter set
recommended in the {\sc dolphot}/ACS User's Guide.

\begin{figure} \includegraphics[width=8cm,angle=0]{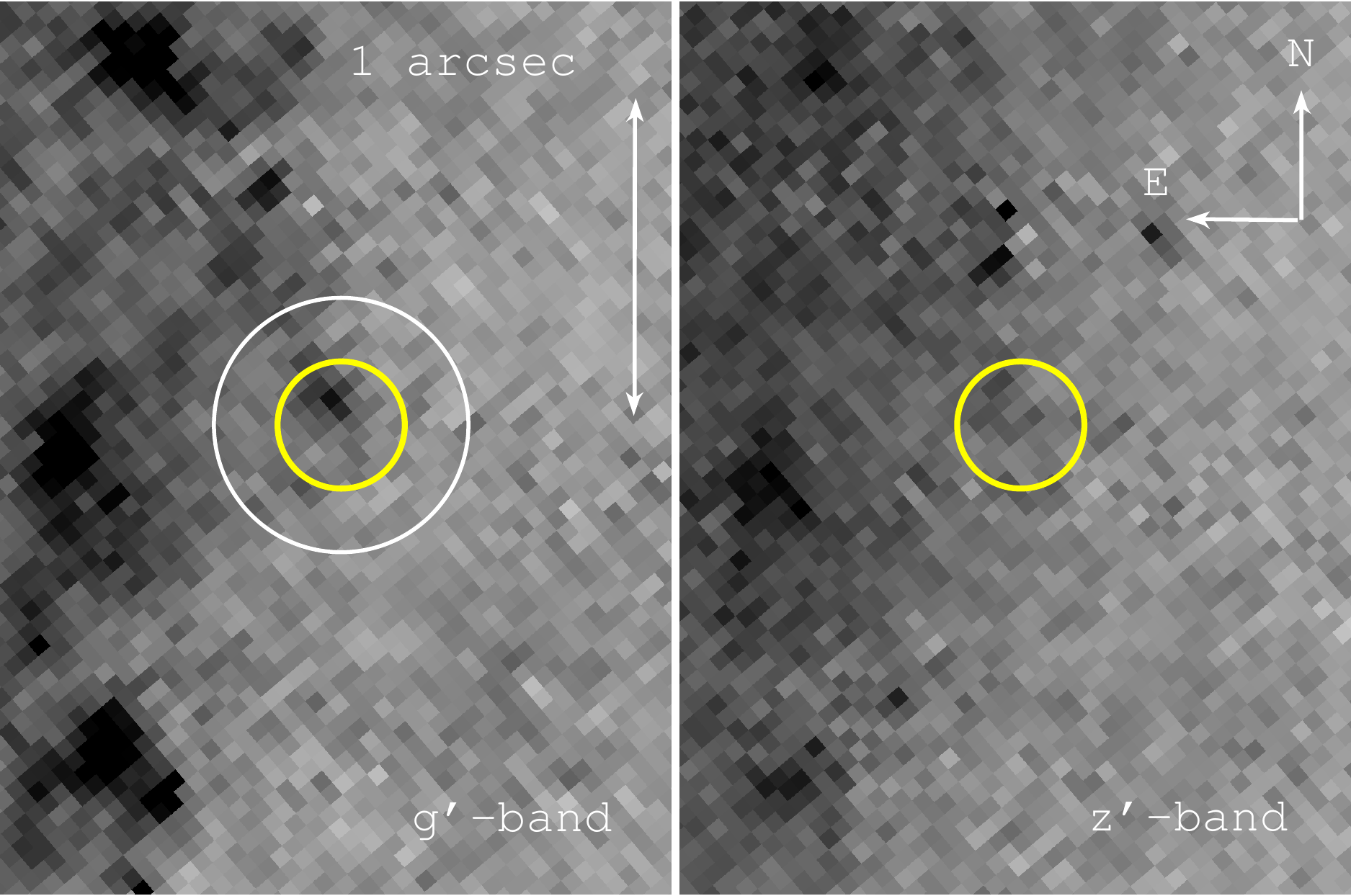} \caption{Zoom-in
of the HST/ACS $g^\prime$-band ({\it left panel}) and $z^\prime$-band ({\it
right panel}) image revealing the (cluster of) stars in the \chan\, error circle
present in the $g^\prime$-band whereas this source is not clearly detectable in
the $z^\prime$-band. The white, smaller circle (yellow in the colour version) is at the \chan\, position and the radius of
0.2\arcsec\, is equal to the overall  astrometric error in the position of the
X-ray source. The larger, white, circle in the $g^\prime$-band image represents the 90 per cent confidence region of the position of \src. The figure is made from a single drizzled $g^\prime$-band and the combined drizzled $z^\prime$-band images.}\label{hstzandgima} \end{figure}

For the photometric analysis we exclude the measurements obtained on
one of the 375~s $g^\prime$-band images due to the presence of charge
due to a cosmic ray hit near the {\it Chandra} position. We improved
the absolute astrometric accuracy of the remaining ACS frame using 5 point
sources that are detected in the SDSS-DR7 $r^\prime$-band image. The resultant error on the astrometry of the ACS {\tt drz}
frame is dominated by the astrometric accuracy of the SDSS $r^\prime$
band which is better than
0.1\arcsec\,(\citealt{2003AJ....125.1559P}). We detect a point source
inside the \chan\, error circle in the remaining 375~s image (see
Fig.~\ref{hstzandgima}) with coordinates R.A. (J2000)=12:25:18.65,
Dec. (J2000)=+14:45:45.8 (R.A.=186.327708, Dec.=14.762722 in decimal degrees) and $g^\prime$ magnitude of $26.4 \pm
0.1$~mag in the VEGA magnitude system. The object is classified as
stellar and is well recovered by the {\sc dolphot} photometric
package. There is no detection of this source in the $z^\prime$-band
images (or any other source within the \chan\, error
circle). Following \citet{2009arXiv0912.3302C} we estimate a
3-$\sigma$ upper limit magnitude at the position of the source of
$z^\prime >$25.7.  This yields $g'-z' \approxlt 0.7$ for the optical
counterpart.

The source is not detected in the \chan\, error circle in the WFPC2
F300W images either. We followed \citet{2009arXiv0912.3302C} using the
latest charge transfer inefficiency corrections parameterized by
\citet{2009PASP..121..655D} to calculate a 3-$\sigma$ upper limit
magnitude of $>22.56$ in the VEGA magnitude system.

\section{Discussion}

We have located a peculiar X-ray source with a position 3.6\arcsec\,
off-nuclear from an SDSS DR7 z=0.0447 galaxy (see Fig.~\ref{ima}
and \ref{hstima}). The 3.6\arcsec\, corresponds to 3.2 kpc
at the distance of the galaxy. The presence of another X-ray source in
the same observation coinciding with a bright optical point source,
allows us to register the \chan\, observation, reducing the
astrometric uncertainty in the position of the off-nuclear
AGN-candidate to less than 0.2\arcsec, making the 3.6\arcsec\, offset
highly significant. The 0.3--8 keV X-ray flux of this source is
5.4$\times 10^{-14}$ \flx\, and its 0.3--8 keV (0.5--10 keV)
luminosity is 2.2$\times 10^{41}$ \lum\, (2.7$\times 10^{41}$ \lum\,)
given the redshift of the galaxy (z=0.0447, d=182.6 Mpc). We find a
candidate optical counterpart in archival HST/ACS g$^\prime$--band
observations of the field containing the galaxy obtained on June 16,
2003. The observed magnitude of g$^\prime=26.4 \pm 0.1$ corresponds to an
absolute magnitude of -10.1 taking a foreground extinction
from our Galaxy of ${\rm A_{g^\prime}=0.18}$ magnitudes into account. These
findings make this source an unusually bright ULX, a very bright
supernova, a recoiling black hole or a background AGN (with higher
luminosity). We discuss these possibilities below.

If this X-ray source is due to a supernova, the 0.5--10 keV X-ray
luminosity is among the highest measured for supernovae. The X-ray
brightest supernovae are of type IIn, this implies that the maximum
optical g$^\prime$-band magnitude was $\approx$17.3 (with mean absolute
blue magnitude of -19.15 type IIn are also optically among the
brightest supernovae; ~\citealt{2002AJ....123..745R}). Now, more
than 6 years after the initial HST/ACS observations, the source
magnitude will have changed considerably. The exact way and amount are
difficult to predict as many scenarios are possible since the optical
lightcurves of type~IIn supernovae are heterogeneous and we do not
know the explosion date.  The supernova could have exploded in between
the serendipitous HST/ACS and the X-ray observation or before the
HST/ACS observation. The X-ray luminosity of the X-ray bright type~IIn
supernovae peaks around 400-1000 days after the explosion making it
impossible to discriminate between the different scenarios on the
basis of the single epoch X-ray and optical observations. The
g$^\prime - z^\prime \approxlt 0.7$ colour of the optical counterpart
candidate is blue for a type IIn supernova origin of the
counterpart (cf.~\citealt{2008PZ.....28....6T} for typical type IIn
colours). If the supernova occurred in between the HST/ACS and the
\chan\, observation the HST/ACS images are of the supernova progenitor
star (cluster).  The SDSS imaging data of this galaxy was taken a few
weeks before the HST data and thus cannot help us deciding between
these scenarios.

In the ULX case the majority of known optical counterparts have blue
colours and they are often embedded in an ionized optical nebula
(\citealt{2008A&A...486..151G} and references therein). In addition,
the counterparts are usually part of small young star clusters or OB
associations (e.g.~\citealt{2002ApJ...577..726Z}). The FWHM of the ACS
image at the distance of the galaxy corresponds to more than 100 pc
hence star clusters would appear as unresolved point sources. 
The ULX counterparts currently known would be below the detection
threshold of the HST/ACS images as their absolute optical magnitude
${\rm M_V}$ is between $-4$ and $-9$
(cf.~\citealt{2008MNRAS.387...73R}) while the source present in the
HST/ACS observation has an absolute $g^\prime$ magnitude of $-10.1$. 

This absolute magnitude is similar to the absolute magnitude of the
counterpart of ESO~243-49 proposed by \citet{2009arXiv0910.1356S}.
The blue colours of the candidate counterpart to \src\, make it hard
to reconcile \src\,with the scenario of an IMBH in a Globular Cluster
(cf.~the colours of Globular Clusters in
\citealt{1996yCat.7195....0H}). Comparing the limit on the $g^\prime -
z^\prime <0.7$ colour with that calculated using the Padua evolution tracks (\citealt{2008A&A...482..883M}; \citealt{2008PASP..120..583G}) we find that the blue colour is consistent with an IMBH in a young, massive, star
cluster.

In the recoiling black hole case the optical counterpart is probably
related to the accretion disc, broad line region and the nuclear
cluster that are retained. The absolute magnitude of the nuclear
region in AGNs varies by more than one order of magnitude, but the
faintest absolute magnitudes are ${\rm M_B}=-9.8$ (NGC~4395;
\citealt{1989ApJ...342L..11F}) and $-11.6$ for NGC~3031/M81 reported
by \citet{1997ApJS..112..391H} in line with our finding. At an
off-nuclear distance of 3.2 kpc, a recoil velocity of $\sim$300 km
s$^{-1}$ is implied if the lifetime of the AGN activity is limited to
$\sim10^7$ years. The hard X-ray spectral power-law index of
0.9$\pm0.3$ that we find implies that there is additional extinction
beyond the Galactic value if the intrinsic source spectrum is
compatible with that of an AGN.

One has to worry about chance alignments on the sky. At an 0.5--10 keV
flux of $5.4\times 10^{-14}$ \flx\,the chance to find a background AGN
in a circle with radius of 3.6\arcsec\, is very small. From the 0.5--10
keV $\log$N -- $\log$S relation found by \citet{2008A&A...492...51M}
there are approximately 76 AGN per square degree.  Using this we
derive that in a circle with radius of 3.6\arcsec\, on average
2$\times10^{-4}$ AGN is found. However, we searched 17 X-ray sources
in galaxies with fluxes equal to or above 5$\times 10^{-14}$\flx, so
considering this number of trials, leads to a probability of 4$\times
10^{-3}$ that the source is a background AGN. 

The X-ray spectrum of the source argues against a background AGN. The
hard X-ray spectrum of the source together with an 0.5-2 keV flux of
$9\pm3\times 10^{-15}$\flx~make that the source falls well below the
trend between power-law index and source 0.5-2 keV flux found by
\citet{2005A&A...444...79M} in their study of the Lockman Hole IV (see
their figure 4). The X-ray to optical flux ratio in the
$g^\prime$-band is 87 and in the $z^\prime$-band this is larger than
84, whereas most \chan\, selected AGNs have values in the $R$-band
below 10 (e.g.~\citealt{2003AJ....126..632B};
\citealt{2009ApJS..180..102L}). If the optical counterpart is
unrelated to the X-ray source the X-ray to optical flux ratio is
larger still. However, if there is additional absorption local to the
AGN, the X-ray to optical flux ratio has been observed to be larger
than 90 (\citealt{2005MNRAS.358..693C}), although there are very few
if any of these sources known at the X-ray flux level of
\src~(cf.~\citealt{2005A&A...444...79M}). If the optical source is
related to the X-ray source but due to a background AGN shining
through the disc of the foreground galaxy that adds a column of
5$\times 10^{21} {\rm cm^{-2}}$, the $g^\prime-z^\prime$ colour limit
would become $<-1.3$ which is uncharacteristically blue for an AGN
(cf.~the $g^\prime-z^\prime$ colours of ROSAT - SDSS AGN presented in
\citealt{2007AJ....133..313A} have a mean $g^\prime-z^\prime$ colour
of $\approx 0.7\pm 0.5$).
 
We conclude that \src\, is a strong candidate for a recoiling SMBH, a
bright ULX with a bright optical counterpart or a very blue type IIn
supernova.

\section*{Acknowledgments} 

\noindent PGJ acknowledges support from a
VIDI grant from the Netherlands Organisation for Scientific Research.
PGJ and MH acknowledge Cees Bassa for help with the bachelor thesis
project of MH. We thank Andrew Dolphin for his help with {\sc dolphot}. We thank the referees for their comments which helped improve the Manuscript. This research has made use of data obtained from the
Chandra Source Catalog, provided by the Chandra X-ray Center (CXC) as
part of the Chandra Data Archive (ADS/Sa.CXO\#CSC). Based on observations made with the Hubble Space Telescope obtained from the ESO/ST-ECF Science Archive Facility.

\end{document}